\documentclass[10pt]{article}
\usepackage{amsmath,amssymb,amsfonts}
\usepackage{units,hyperref,fancyhdr}
\usepackage{bbold,mathpazo}
\usepackage{graphicx}
\usepackage{dcolumn}
\usepackage{bm,bbm,mathtools,esvect}
\usepackage{slashed}
\usepackage{fullpage}
\title{\bf SELF-INTERACTING  QUANTUM  PARTICLES \\  \vspace{5mm} }
\author{{\sf SERGIO GIARDINO\footnote{\tt sergio.giardino@ufrgs.br}}\\
\\
\small \it Departamento de Matem\'atica Pura e Aplicada \\
\small \it Universidade Federal do Rio Grande do Sul (UFRGS)\\
\small \it Caixa Postal 15080, 91501-970  Porto Alegre RS \\
\small \it Brazil}

\begin{document}
\date{}
\maketitle

\begin{abstract}
\noindent The real Hilbert space formalism developed within the quaternionic quantum mechanics ($\mathbbm H$QM) is fully applied to the simple model of the autonomous particle. This framework permits novel insights within the usual description of the complex autonomous particle, particularly concerning the energy of a non-stationary motion. Through the appraisal of the physical role played by a fully quaternionic scalar potential, a original self-interaction within the quaternionic autonomous particle has been determined as well. Scattering processes are considered to illustrate these novel features.

\vspace{2mm}

\noindent {\bf keywords:} quantum mechanics; formalism; Scattering theory; other topics in mathematical methods in physics

\vspace{1mm}

\noindent {\bf pacs numbers:} 03.65.-w; 03.65.Ca; 03.65.Nk; 02.90.+p. 
\end{abstract}

\vspace{3cm}

\hrule
{\parskip - 0.3mm \footnotesize{\tableofcontents}}
\vspace{1cm}
\hrule

\pagebreak 

\section{INTRODUCTION\label{I}}%

Imaginary components of complex scalar potentials in the Hamiltonian operator describes non-stationary quantum processes, like the inelastic scattering ({\it cf.} \cite{Schiff:1968qmq,Muga:2004zz} Section 20). However, in quaternionic quantum mechanics ($\mathbbm H$QM), where the imaginary component of a quaternionic scalar potential comprises three imaginary units, the physical meaning of each component is not understood, and this article aims to clarify this essential point taking benefit of one of the simplest solutions of quantum mechanics: the autonomous free particle.

Inevitably, the simplicity was the criterion used to elect the autonomous particle as the correct model to investigate the physical properties of the fully quaternionc scalar potential. Further, one should recollect the simplest solutions as the most important results of every physical theory, illuminating the most fundamental properties, and constituting the bare elements to fabricate the sophisticated solutions needed by more complicated physical situations. Additionally, either when a  modification is introduced in a theory, or when an entire novel theory is formulated, primary solutions are the ideal way to test these innovative ideas. These elementary principles invariably hold also in case of quantum mechanics, and the quantum autonomous particle will be deployed here as a theoretical device to investigate basic features of the Hamiltonian operator and of the  wave equation within the theoretical framework of the real Hilbert space.

Before going into the details of the calculations, and also remembering that quaterionic theories also have experimental interest \cite{Walter:2017prw,Adler:2016ciu,Procopio:2017vwa}, one notices the quaternionic quantum mechanics ($\mathbbm H$QM) to be a mathematical formalism in which quantum mechanics is composed in terms of the four dimensional generalized complex numbers known as quaternions ($\mathbbm H$). Strictly speaking, in $\mathbbm H$QM the quaternions replace the complex numbers ($\mathbbm C$) that sustain the usual theory ($\mathbbm C$QM). In the same manner as quaternions generalize complexes, one can expect that $\mathbbm H$QM mathematically generalizes quantum mechanics. The query whether quantum mechanics admits some mathematical generalization is the main motivation for $\mathbbm H$QM. From a physical standpoint, one inquires whether the  the current form of quantum mechanics is mathematically adequate to understand the reasons why several fundamental theories as string theory and general relativity resist to quantization. If the generality of $\mathbbm C$QM is not sufficient, these questions will remain unsolved until the replacement of the theory. At the present time, $\mathbbm H$QM is still only a candidate to such possible generalized quantum theory.

However, there are several applications of quaternions in quantum mechanics, and not all of them intents to be a generalization. Since the quaternionic generalization of quantum mechanics is not straightforward, there are two main theoretical proposals to $\mathbbm H$QM. The older one uses the quaternionic Hilbert space, and a more recent proposal uses the real Hilbert space. Also referred as the anti-hermitean $\mathbbm H$QM, the quaternionic Hilbert space proposal requires anti-hermitean Hamiltonian operators, and comprises a vast amount of work contained in a seminal book by Stephen Adler \cite{Adler:1995qqm}. Nonetheless, serious drawbacks plague this theory, first of all the ill-defined classical limit ({\it c.f.} sec. 4.4 of \cite{Adler:1995qqm}), implying the Ehrenfest theorem not to hold. A further serious disadvantage is the highly involved formulation of the anti-hermitian theory, meaning that simple solutions are hard to find, and to interpret. One can quote several examples of such solutions involving themes like scattering \cite{DeLeo:2015hza,Sobhani:2016qdp,Hassanabadi:2017wrt,Hassanabadi:2017jiz,Procopio:2017vwa,Sobhani:2017yee}, operators and potentials \cite{Ducati:2001qo,Nishi:2002qd,Madureira:2006qps,DeLeo:2013xfa}, wave packets \cite{Ducati:2007wp}, quantization methods \cite{Muraleetharan:2014qma,Bolokhov:2017ndw,Sabadini:2017wha}, bound states \cite{DeLeo:2005bs,Giardino:2015iia}, perturbation theory \cite{DeLeo:2019bcw}, high dimensional physics \cite{Brody:2011mg}, and quantum computing \cite{Dai:2023xxh}. Notwithstanding, a clear and operational interpretation of anti-hermitean $\mathbbm H$QM does not come from them. On the other hand, there are quaternionic applications to $\mathbbm C$QM, where the Hilbert space is still complex, but quaternionic structures appear within various theoretical objects, such as operators and wave functions  \cite{Arbab:2010kr,Graydon:2013sra,Sapa:2020dqm,Rawat:2022xsj,guo:2024sqm,Danielewski:2023acc}, Dirac's monopoles \cite{Soloviev:2016qsx}, quantum states \cite{Steinberg:2020xvf}, angular momenta \cite{deepka:2024nsf}, fermions \cite{Cahay:2019bqp,Cahay:2019pse}, and in the mass concept \cite{Arbab:2022cpe}. These quaternionic application can be classified as mathematical methods of solutions to the usual complex quantum theory, and do not represent any conceptual generalization of quantum mechanics, although they can be useful in specific cases.

On the other hand, several drawbacks of the anti-hermitean approach can be suppressed using the real Hilbert space formalism \cite{Giardino:2018rhs}, where a well-defined classical limit holds \cite{Giardino:2018lem}, and  simple quaternionic systems have been solved, comprising the Aharonov-Bohm effect \cite{Giardino:2016xap}, autonomous particle solutions \cite{Giardino:2017yke,Giardino:2017pqq}, the Virial theorem \cite{Giardino:2019xwm}, the quantum elastic scattering \cite{Giardino:2020ztf,Hasan:2019ipt}, rectangular potentials \cite{Giardino:2020cee}, the harmonic oscillator \cite{Giardino:2021ofo}, spin \cite{Giardino:2023spz}, and generalized imaginary units \cite{Giardino:2023uzp} Quantum relativistic solutions have been also accomplished using the real Hilbert space approach, including the Klein-Gordon equation \cite{Giardino:2021lov}, the Dirac equation \cite{Giardino:2021mjj}, the scalar field \cite{Giardino:2022kxk}, and the Dirac field \cite{Giardino:2022gqn}. An important feature of the real Hilbert space approach to $\mathbbm H$QM is the definition of the expectation value of an arbitrary quantum operator, 
\begin{equation}\label{pa19}
 \big\langle\widehat{\mathcal O}\big\rangle =\frac{1}{2}\int d\bm x\left[\Psi^\dagger \widehat{\mathcal O}\Psi+\Big(\widehat{\mathcal O}\Psi\Big)^\dagger\Psi\right],
\end{equation}
where $\Psi$ is the quaternionic wave function, and $\Psi^\dagger$ is their conjugate. The expectation value (\ref{pa19}) always give real values, and the quantum operator $\widehat{\mathcal O}$ needs not to be hermitian, what represents a crucial attribute of the real Hilbert space approach.

Previous solutions of $\mathbbm H$QM in the real Hilbert space only considered real scalar potentials. In this article, conversely, the intention is to explore more general scalar potentials. The strategy of using  the autonomous particle solution as a way to to deep the understanding of several topics of the usual $\mathbbm C$QM is of course not new, and one can mention thermal wave packets \cite{Bindech:2020zto}, quantum mechanics in curved space \cite{Teixeira:2019icw,Frick:2014zza,Sato:2022tbi}, isospectral Hamiltonians \cite{Angelone:2022eee}, quantum measurement \cite{Gampel:2023jyy}, quantum properties \cite{Bracken_2021}, and also the quantum Zeno effect \cite{PhysRevA.90.062131}.
Following this idea, one will consider complex and quaternionic autonomous particles, their wave functions, energy conservation, and scattering through a rectangular barrier, a topic that is also of contemporary physical interest \cite{los:2010pat,los:2013exa,Gadella:2016lum}.

\section{COMPLEX PARTICLES\label{P}}%

Autonomous quantum particles are simple and well known solutions of quantum mechanics, and one will consider them in this section in order to be a model for the quaternionic self-interacting particle. Notwithstanding, the real Hilbert space approach reveal interesting features of this solution that cannot be achieved within the usual complex Hilbert space approach. However, a higher degree of mathematical generality than that usually found in textbooks is required to the complex solution to fulfill the requirements of a suitable prototype to the quaternionic solutions, particularly to establish the criteria of stationary quaternionic states. 
To establish this complex template, one thereby recalls that to a quantum particle of mass $m$ corresponds a wave function $\psi$ that solves the Schr\"odinger equation 
\begin{equation}\label{pa01}
 i\hbar\frac{\partial \psi}{\partial t}=\left(-\frac{\hbar^2}{2m}\nabla^2+V\right)\psi,
\end{equation}
where the constant complex scalar potential
\begin{equation}
 V=V_0+iV_1,
\end{equation}
holds everywhere in space, and $V_0$ and $V_1$ are real constants. The naive solution accordingly is
\begin{equation}\label{pa17}
 \psi(\bm x,\,t)=\phi(\bm x)\exp\left[-\frac{E}{\hbar}t\right],
\end{equation}
where $\bm x$ is the position vector and $E$ is the complex constant
\begin{equation}\label{pa39}
 E=E_0+E_1i.
\end{equation}
Subsequently, the time independent equation reads
\begin{equation}\label{pa26}
 \nabla^2\phi=\frac{2m}{\hbar^2}\Big[V_0-E_1+i\big(V_1+E_0\big)\Big]\phi,
\end{equation}
whose solution comprises
\begin{equation}\label{pa18}
 \phi=A\exp\big[\bm{K\cdot x}\big],\qquad \mbox{where} \qquad \bm K=\bm K_0+\bm K_1 i,
\end{equation}
where the amplitude $A$ is a complex constant, as well as $\bm K_0$ and $\bm K_1$ are real constant vectors. Of course, there is a second solution, where a flipped signal holds in the argument of the exponential, so that 
$\phi=A\exp\big[-\bm{K\cdot x}\big]$, and this solution corresponds to a free particle of opposite direction, as we will see. To determine the conditions involving $V$ and $E$ that generate a stationary motion, substituting (\ref{pa18}) in (\ref{pa26}) renders
\begin{equation}\label{pa27}
\|\bm K_0\|^2-\|\bm K_1\|^2=\frac{2m}{\hbar^2}\Big(V_0-E_1\Big),\qquad \mbox{and}\qquad 2\bm K_0\bm{\cdot K}_1=\frac{2m}{\hbar^2}\Big(V_1+E_0\Big).
\end{equation}
Assuming 
\begin{equation}\label{uni18}
 \bm K_0\bm{\cdot K}_1=\|\bm K_0\|\|\bm K_1\|\cos\Omega_0,
\end{equation}
where $\Omega_0$ is a phase angle, the above result is the farthest point to be reached in various dimensions. In one dimension, where constants replace the vectors and a multiplication replaces the scalar product (\ref{pa27}), the real components of $K$ are as follows
\begin{equation}\label{pa33}
\| \bm K_0\|^2=\frac{m}{\hbar^2}\left(V_0-E_1+\sqrt{\big(E_1-V_0\big)^2+\left(\frac{V_1+E_0}{\cos\Omega_0}\right)^2}\,\right)
\end{equation}
and 
\begin{equation}\label{pa47}
 \|\bm K_1\|^2=\frac{m}{\hbar^2}\left(E_1-V_0+\sqrt{\big(E_1-V_0\big)^2+\left(\frac{V_1+E_0}{\cos\Omega_0}\right)^2}\,\right),
\end{equation}
where $\cos\Omega_0\neq 0$ is implicit. The reality of $K_0$ and $K_1$ eliminated the possibility of a minus signal before the square roots. Before determining the relation between $E$ and  $V$ that enables stationary solutions, one can study their physical character in terms of expectation  values, and of the conservation of the probability. 

\paragraph{CONSERVATION LAWS} First of all, one defines the energy, and the linear momentum operators as
\begin{equation}\label{pa30}
 \widehat E=i\hbar\frac{\partial}{\partial t},\qquad  \mbox{and}\qquad
 \widehat p=-i\hbar\frac{\partial}{\partial x}.
\end{equation}
Thus, the parameters of the solution can be interpreted in terms of probability density after recalling that the probability scalar density $\rho$, the probability current vector $\bm J$, and the probability scalar source $g$, satisfy the continuity equation \cite{Giardino:2018rhs},
\begin{equation}\label{pa29}
\frac{\partial\rho}{\partial t}+\bm{\nabla\cdot J}=g,
\end{equation}
where each term accordingly reads
\begin{equation}
 \rho=\psi\psi^\dagger,\qquad \bm J=\frac{1}{2m}\left[\psi\Big(\widehat{\bm p}\psi\Big)^\dagger+\Big(\widehat{\bm p}\psi\Big)\psi^\dagger\right],\qquad \mbox{and}\qquad
 g=\frac{1}{\hbar}\rho\Big(\overline V i-iV\Big).
\end{equation}
By reason of (\ref{pa17}) and (\ref{pa18}), one obtains
\begin{equation}\label{pa36}
 \psi(x,\,t)=A\exp\left[Kx-\frac{E}{\hbar}t\right],
\end{equation}
as well as
\begin{equation}\label{pa28}
 \rho=|A|^2\exp\left[2K_0x-\frac{2E_0}{\hbar}t\right],\qquad J=\frac{\hbar K_1}{m}\rho\qquad g=\frac{2V_1}{\hbar}\rho.
\end{equation}
As expected, the imaginary component $V_1$ of the scalar potential $V$ is associated to the source of probability, and accordingly to non-stationary processes. The probability density either increases or decreases because of the real components of $E$ and $K$, confirming that $E_0,\,K_0$ and $V_1$ are responsible by non-stationary processes. Finally,  (\ref{pa28}) into the continuity equation (\ref{pa29}), one recovers the imaginary component of (\ref{pa27}), indicating that the physical information concerning the conservation of the probability is also contained in the wave equation.
Using the operators (\ref{pa30}), and the definition of the expectation value (\ref{pa19}), one obtains
\begin{eqnarray}
\nonumber&& \left\langle \widehat E\right\rangle=E_1\int\!\!\rho \,dx,\\
\nonumber && \Big\langle \widehat{\bm p}\Big\rangle=\hbar \bm K_1 \int\!\!\rho\, dx,\\
\label{pa32}&& \left\langle \|\widehat{\bm p}\|^2\right\rangle=\hbar^2\Big(\|\bm K_1\|^2-\|\bm K_0\|^2\Big) \int\!\!\rho\, dx,\\
\nonumber&&\left\langle \widehat V\right\rangle=V_0\int\!\!\rho\, dx.
\end{eqnarray}
In the first instance, one observes that the above real expectation values cannot be obtained in the standard $\mathbbm C$QM, because the imaginary component cannot be eliminated in the usual definition of the inner product, and thus the real quantities (\ref{pa32}) are a particular attribute of the real Hilbert space expectation value (\ref{pa19}). In other words, the usual complex Hilbert space result is recovered only if $E_0=K_0=0$, as expected, but otherwise the $\mathbbm C$QM formalism is unsuitable.

Likewise the complex Hilbert space case, the wave function does not admit normalization if $K_0=0$, but $E_0\neq 0$ imposes the expectation values either to increase or to decrease according to an identical rate in time, determined  by the exponential $\exp[-2E_0 t/\hbar]$, a result that cannot be obtained within the complex Hilbert space  quantum mechanics. Additionally, one observes that $K_1$ alone determines the direction $\bm p$, and $K_0\neq 0$ only contributes to the integral of $\rho$, both in accordance to $\mathbbm C$QM.

A notable feature of (\ref{pa32}) concerns the conservation of the energy
\begin{equation}\label{pa31}
 \left\langle \widehat E\right\rangle=\frac{1}{2m}\left\langle \widehat p^2\right\rangle+\left\langle \widehat V\right\rangle,
\end{equation}
where the dependence on the probability density factors out, thus recovering the relation involving the parameters contained in the real component of (\ref{pa27}). One can understand (\ref{pa27}) as a global property, while the dependence on time within the probability density $\rho$ determines the precise situation of this relation in every instant of the elapsed time, and one can then identify (\ref{pa31}) with a local property as well. In simple words, even if the motion can is evanescent, or forced, it preserves the energy relation that contains the mechanical character of the system. 

Remarkably, the conservation of the energy holds even in case of negative or null squared linear momentum $p^2$, what can be obtained  if $|K_1|^2<|K_1|^2$ and $E_1<V_0$, determining the  motion to exhibit a non-stationary quantum character. This phenomenon cannot be explained neither in the complex Hilbert space formalism, nor in terms of classical mechanics, although negative quantum energies have already been considered elsewhere \cite{Pavsic:2020aqi,Geszti:2024ddw} within a non-linear context. 

\paragraph{STATIONARY STATES} The next task is to examine the conditions to have stationary states of the autonomous particle. Requiring the parameters $E$, and $V$ to be chosen from beginning, and $K$ to be determined in (\ref{pa33}-\ref{pa47}),  the stationary motion along the time variable is simply a choice, and thus
\begin{equation}\label{pa34}
 E_0=0
\end{equation}
is the only requirement to have a time stationary particle. In terms of the space variable, there is also only one possible stationary state, where
\begin{equation}
 K_0=0
\end{equation}
determined by condition (\ref{pa33}) is such that. 
\begin{equation}\label{pa35}
 V_0<E_1\qquad \mbox{and}\qquad V_1+E_0=0.
\end{equation}
Combining (\ref{pa34}) and (\ref{pa35}) one arrives at
\begin{equation}
 V_0-E_1<0\qquad \mbox{and}\qquad V_1=E_0=0,
\end{equation}
to be the condition of the autonomous particle, an expected result. Of course, there are various possibilities for non-stationary autonomous particles, where $E$ and $K$ are not pure imaginary. However, one can state that nonzero $E_0$ and $V_1$ always generate non-stationary solutions.

\paragraph{SCATTERING} Finally, one can consider the one-dimensional scattering of an autonomous particle according to the complex potential
\begin{equation}\label{pa40}
 V=\left\{
 \begin{array}{lll}
 V_I & \mbox{if} & x<0\\
 V_{II} & \mbox{if} & x\geq 0,
 \end{array}
 \right.
\end{equation}
where $V_I$ and $V_{II}$ are constant complex potentials. Conforming to (\ref{pa36}), the wave function describing the scattering of a particle that travels in the region submitted to potential $V_I$, and gets into the region governed by potential $V_{II}$ at the point $x=0$ reads
\begin{equation}\label{pa53}
 \psi=\left\{ 
 \begin{array}{lll}
  \psi_I=\Big(\exp\big[K_Ix\big]+R\exp\big[- K_Ix\big]\Big)\exp\left[-\frac{E_I}{\hbar}t\right] & \mbox{if} & x<0\\
  \psi_{II}=T\exp\left[K_{II}x-\frac{E_{II}}{\hbar}t\right] & \mbox{if} & x\geq 0,
 \end{array}
 \right.
\end{equation}
where $E_I,\,E_{II},\,K_I,\,$ and $K_{II}$ are complex constants, as well as $R$ and $T$. The wave function  and its first spatial derivative at $x=0$ will be required to satisfy the conditions
\begin{equation}\label{pa51}
 \big|\psi_I(0,\,t)\big|^2=\big|\psi_{II}(0,\,t)\big|^2, \qquad \mbox{and}\qquad \big|\psi'_I(0,\,t)\big|^2=\big|\psi'_{II}(0,\,t)\big|^2,
\end{equation}
or equivalently
\begin{equation}\label{pa52}
 \psi_I(0,\,t)=\psi_{II}(0,\,t)\exp[i\varphi_0], \qquad \mbox{and}\qquad \psi'_I(0,\,t)=\psi'_{II}(0,\,t)\exp[i\xi_0].
\end{equation}
The above conditions are more general than the usual continuity condition, which is recovered within the limit $\varphi_0=\xi_0=0.$ Conversely, (\ref{pa51}) has a clear physical interpretation in terms of the conservation of the number of particles at the boundary, but the usual continuity condition is unnecessarily tighter. Therefore, (\ref{pa52}) seems more suitable to allow further physical phenomena to appear. The energy of the particle does  not change after crossing the border between the regions, and therefore
\begin{equation}
 E_I=E_{II}=E,
\end{equation}
where of course (\ref{pa39}) holds. Using (\ref{pa53}) and (\ref{pa52}), one immediately achieves
\begin{equation}\label{pa37}
|R|^2=\frac{\big|K_Ie^{i\varphi_0}-K_{II}e^{i\xi_0}\big|^2}{\big|K_Ie^{i\varphi_0}+K_{II}e^{i\xi_0}\big|^2}\qquad
\mbox{and}\qquad |T|^2=\frac{4|K_I|^2}{\big|K_Ie^{i\varphi_0}+K_{II}e^{i\xi_0}\big|^2}
\end{equation}
and also
\begin{equation}\label{pa38}
 |R|^2+|T|^2=1+u,
\end{equation}
where $u$ is a factor that indicates the conservation of the particles within the scattering process, so that
\begin{equation}\label{pa54}
 u=2\frac{K_Ie^{i\varphi_0}\Big(\overline K_Ie^{-i\varphi_0}-\overline K_{II}e^{-i\xi_0}\Big)+\overline K_I e^{-i\varphi_0}\Big(K_Ie^{i\varphi_0}-K_{II}e^{i\xi_0}\Big)}{\big|K_Ie^{i\varphi_0}+K_{II}e^{i\xi_0}\big|^2}.
\end{equation}
In other words, if $u=0$ the transition between the regions does not involve neither creation nor annihilation of particles, and this condition conservation requires the condition
\begin{equation}
 K_Ie^{i\varphi_0}-K_{II}e^{i\xi_0}=0.
\end{equation}
Of course, the process in non-consevative even in the usual complex case, when $K_I$ as well as $K_{II}$ are pure imaginary, and is not associated to stationary processes. However, the conservative phenomenona contained in condition (\ref{pa52}) may include the evanescence of the scattered particle, a novel and interesting case for future directions of research.

A final comment can be obtained using (\ref{pa27}) and their interpretation in terms of the conservation of the energy. Because the energy parameters are identical in both of the regions, the equality of $E_1$ in both of the regions of the scattering process immediately generates
\begin{equation}
 \frac{1}{2m}\Delta p^2+\Delta\mathfrak{Re}[V]=0,
\end{equation}
demonstrating that the increase in the potential inside one the regions means the increase of the kinetic energy inside the other of the regions. These results summarize what the most important differences between the description of an autonomous particle within the real Hilbert space formalism and the usual complex Hilbert space. The presented outcomes generate a clear advantage because it unifies stationary and non-stationary states within a single description.  The complex autonomous particle is also the prototype of to be referred in the quaternionic cases to be considered in the next sections.

\section{QUATERNIONIC PARTICLES I}

The quaternionic autonomous particles in the real Hilbert space have already been described in terms of real scalar potentials \cite{Giardino:2017yke,Giardino:2017pqq}, and the complete quaternionic scalar potentials will be considered in this article. One recalls the wave function $\Psi$ evaluated over quaternions to be
\begin{equation}\label{pa02}
 \Psi=\psi_0+\psi_1 j,
\end{equation}
where $\psi_0$ and $\psi_1$ are complex functions, and $j$ an imaginary unit. The basic facts concerning quaternions can be obtained from various sources \cite{Ward:1997qcn,Morais:2014rqc,Ebbinghaus:1990zah}, and will not be provided here. Nevertheless, one must emphasize two consequence of the adoption of quaternions: the bigger number of degrees of freedom, represented by the additional complex function $\psi_1$ in (\ref{pa02}), and the non-commutativity, whose initial consequence is the existence of two possible wave equations that generalize the complex Schr\"odinger equation. Due to the fact that
\begin{equation}
 \Psi i\neq i\Psi,
\end{equation}
the multiplication order between the time derivative of the wave function, and the imaginary unit $i$  generate two viable wave equations. The first alternative reads
\begin{equation}\label{pa04}
i \hbar\frac{\partial\Psi}{\partial t}=\widehat{\mathcal H}\Psi,
\end{equation}
and the second possibility, where the imaginary unit multiplies the right hand side of the time derivative, is considered in the next section. The Hamiltonian operator $\widehat{\mathcal H}$, which is exactly the same in both of the cases, defines
\begin{equation}\label{pa03}
 \widehat{\mathcal H}=-\frac{\hbar^2}{2m}\nabla^2+U,
\end{equation}
and it differs from the Hamiltonian of the complex Schroedinger equation (\ref{pa01}), because the scalar potential $U$ is quaternionic, so that
\begin{equation}\label{pa25}
 U=U_0+U_1j,
\end{equation}
and the complex components of $U$ comprise
\begin{equation}\label{pa15}
 U_0=V_0+V_1 i\qquad \mbox{and}\qquad U_1=W_0+W_1 i,
\end{equation}
where   $V_0,\,V_1,\,W_0$ and $W_1$ are of course real. The generalization of the complex case considered in the previous section requires the real components of $U$ to be constant. The wave function (\ref{pa02}) and the constant potential (\ref{pa25}) substituted in the wave equation (\ref{pa04}) generate a pair of complex equations, so that
\begin{eqnarray}
\label{pa06} && i\hbar\frac{\partial\psi_0}{\partial t}=-\frac{\hbar^2}{2m}\nabla^2\psi_0+U_0\psi_0-U_1\psi_1^\dagger\\
\label{pa07} && i\hbar\frac{\partial\psi_1}{\partial t}=-\frac{\hbar^2}{2m}\nabla^2\psi_1+U_0\psi_1+U_1\psi_0^\dagger.
\end{eqnarray}
Equation (\ref{pa06}) comes from the complex component of the wave equation, and (\ref{pa07}) accordingly comes from the pure quaternionic component of the same equation, and $\psi^\dagger$ represents the complex conjugate of the wave function $\psi$. This pair of differential equations depicts the increase of the degrees of freedom generated by the quaternionic generalization of the Schr\"odinger equation. Besides, the pair of complex equations (\ref{pa06}-\ref{pa07}) reveals that $U_1$ generate the coupling between $\psi_0$ and $\psi_1$, and one must stress that this coupling produce a self-interaction within the particle, a undeniably novel feature of quantum theory. Inevitably, the pure quaternionic potential imposes the separation between the self-interacting and non-self-interacting cases, and one has to consider them separately. One also remarks the similarity between (\ref{pa06}-\ref{pa07}) and the recently proposed non-linear quantum model \cite{Chodos:2022ksw,Chodos:2024xby}. The relation between these theories is an interesting direction for future research.

\paragraph{THE NON-SELF-INTERACTING PARTICLE} If $U_1=0$ in (\ref{pa25}), the complex components $\psi_0$ and $\psi_1$ of the quaternionic wave function $\Psi$ are completely independent, and therefore their energies and momenta are free to assume every value. Hence, the quaternionic wave function under a complex potential $V$ in one dimension assumes the form
\begin{equation}\label{pa41}
 \Psi=A\exp\left[\bm{K\cdot x}-\frac{E}{\hbar}t\right]+\mathcal A\exp\left[\bm{ \mathcal K\cdot x}-\frac{\mathcal E}{\hbar}t\right]j,
\end{equation}
where $A,\,\bm K,\,E,\,\mathcal A,\,\bm{\mathcal K}$ and $\mathcal E$ are complex quantities, as discussed in the complex case. Following (\ref{pa32}), one obtains the expectation values
\begin{eqnarray}
\nonumber&& \left\langle \widehat E\right\rangle=E_1\int\!\!\rho \,dx+\mathcal E_1\int\!\!\varrho \,dx,\\
\nonumber && \Big\langle \widehat{\bm p}\Big\rangle=\hbar\bm K_1 \int\!\!\rho\, dx+\hbar \bm{\mathcal K}_1 \int\!\!\varrho\, dx,\\
\label{pa44}&& \left\langle \|\widehat{\bm p}\|^2\right\rangle=\hbar^2\Big(\mathcal \|\bm{K}_1\|^2-\|\bm{K}_0\|^2\Big) \int\!\!\rho\, dx+\hbar^2\Big(\mathcal \|\bm{\mathcal K}_1\|^2-\|\bm{\mathcal K}_0\|^2\Big) \int\!\!\varrho\, dx,\\
\nonumber&&\left\langle \widehat V\right\rangle=V_0\int\!\!\big(\rho+\varrho\big)\, dx,
\end{eqnarray}=
where $\bm{\mathcal K}_0,\,\bm{\mathcal K}_1,\,\mathcal E_0$ and $\mathcal E_1$ are the real components of $\bm{\mathcal K}$ and $\mathcal E$, the density of probability $\rho$ of $\psi_0$ conforms (\ref{pa28}), and $\varrho$ is of course the probability density corresponding to $\psi_1$. The expectation values of the quaternionic non-self-interacting particle are simply the outcome of the sum of the expectation values of each independent complex wave function. However, two constraints involving the complex components can be obtained from the energy conservation (\ref{pa27}), from the continuity equation,  and from the linear independence of $\rho$ and $\varrho$, such as
\begin{equation}\label{pa48}
E_1 -\frac{\hbar^2}{2m}\Big(\|\bm K_1\|^2-\|\bm K_0\|^2\Big)=\mathcal E_1 -\frac{\hbar^2}{2m}\Big(\|\bm{\mathcal K}_1\|^2-\|\bm{\mathcal K}_0\|^2\Big)=V_0
\end{equation}
and also
\begin{equation}\label{pa49}
E_0- \frac{\hbar^2}{m}\bm K_0\bm{\cdot K}_1=\mathcal E_0 - \frac{\hbar^2}{m}\bm{\mathcal K}_0\bm{\cdot\mathcal K}_1=V_1.
\end{equation}
Relations (\ref{pa48}-\ref{pa49}) confirm the physical content of the non-self-interacting quaternionic particle to be contained in the wave equation, and further demonstrate that only a weak constraint can be established between the complex components. Finally, one can consider the scattering of a quaternionic autonomous particle by the complex potential (\ref{pa40}), where the vector components can be managed as real numbers, and the wave function accordingly is
\begin{equation}
 \Psi=\left\{
 \begin{array}{ll}
  \Psi_I=\exp\left[K_Ix-\frac{E_I}{\hbar}t\right]+\exp\left[\mathcal K_I x-\frac{\mathcal E_I}{\hbar}t\right]j+R\left(\exp\left[- K_Ix-\frac{E_I}{\hbar}t\right]+\mathcal A\exp\left[-\mathcal K_I x-\frac{\mathcal E_I}{\hbar}t\right]j\right)\\ \\
  \Psi_{II}=T\Big(\exp\left[K_{II}x-\frac{E_{II}}{\hbar}t\right]+\mathcal B\exp\left[\mathcal K_{II} x-\frac{\mathcal E_{II}}{\hbar}t\right]j\Big),
 \end{array}
 \right.
\end{equation}
where $R,\,T,\,\mathcal A$ and $\mathcal B$ are complex constants. The continuity at $x=0$ of the wave function establishes the identity of the complex energy constants, so that
\begin{equation}\label{pa16}
 E_I=E_{II}\qquad \mbox{and}\qquad \mathcal E_I=\mathcal E_{II}.
\end{equation}
Likewise, the continuity of the space derivative at $x=0$ determines identical system of equations for $R$ and $T$ in terms of $K$, and consequently (\ref{pa37}-\ref{pa38}) hold for the first complex component of the wave function. The solution set equally holds for the complex component component coming from the pure quaternionic component in terms of the correspondence
\begin{equation}
 R\to R\mathcal A,\qquad T\to T\mathcal B,\qquad K\to \mathcal K,
\end{equation}
so that
\begin{equation}\label{pa42}
 |R\mathcal A|^2+|T\mathcal B|^2=1+v,
\end{equation}
where
\begin{equation}
 v=2\frac{\mathcal K_I\Big(\overline{\mathcal K}_Ie^{-i\varphi_0}-\overline{\mathcal K}_{II}e^{-i\xi_0}\Big)+\overline{\mathcal K}_I\Big(\mathcal K_I e^{i\varphi_0}-\mathcal K_{II}e^{i\xi_0}\Big)}{\big|\mathcal K_Ie^{i\varphi_0}+\mathcal K_{II}e^{i\xi_0}\big|^2}.
\end{equation}
Expectedly, each complex component behaves as an independent particles, and the case $\mathcal A=\mathcal B=1$ implies that both of the complex behave identically, and thus the phenomenology of the complex case is recovered, as desired. If $\mathcal A\neq \mathcal B$, from (\ref{pa38}) and (\ref{pa42}) one obtains
\begin{equation}
 |R|^2=\frac{1-|\mathcal B|^2}{|\mathcal A|^2-|\mathcal B|^2}+\frac{v-u|\mathcal B|^2}{|\mathcal A|^2-|\mathcal B|^2},
\end{equation}
and
\begin{equation}
 |T|^2=\frac{|\mathcal A|^2-1}{|\mathcal A|^2-|\mathcal B|^2}+\frac{u|\mathcal A|^2-v}{|\mathcal A|^2-|\mathcal B|^2}.
\end{equation}
Therefore, the conservation relation (\ref{pa38}) of the complex scattering holds, and the difference between the complex case and the non-interacting quaternionc case concerns exclusively to the constraints (\ref{pa48}-\ref{pa49}), indicating that the parameters of the solutions are not independent, although the difference to the complex case seems to be physically irrelevant.

\paragraph{THE SELF-INTERACTING PARTICLE}
For the purpose of solving the coupled case of (\ref{pa06}-\ref{pa07}), where $U_1\neq 0$, one separates time and spatial variables, such as
\begin{equation}\label{pa08}
 \psi_0=\phi_0(\bm x)\exp\left[-\frac{\varepsilon_0}{\hbar}t\right],\qquad \mbox{and}\qquad
 \psi_1=\phi_1(\bm x)\exp\left[-\frac{\varepsilon_1}{\hbar}t\right],
\end{equation}
where $\phi_0$ and $\phi_1$ are complex functions, and $\varepsilon_0$ and $\varepsilon_1$ are complex constants. Nonetheless, the variables can be separated only in the case of the complex energy parameters related by a conjugation relation, so that
\begin{equation}\label{pa09}
\varepsilon_0=\overline{\varepsilon}_1=E,
\end{equation}
where $E$ conforms to (\ref{pa39}). Condition (\ref{pa09}) constraints the energy parameters of the complex components of a self-interacting quaternionic particle in a way that is not observed within the previous non-self-interaction case. Using (\ref{pa08}-\ref{pa09}) in the system of equations (\ref{pa06}-\ref{pa07}), one obtains
\begin{eqnarray}
\nonumber &&  \frac{\hbar^2}{2m}\nabla^2\phi_0=\big(U_0+iE\big)\phi_0-U_1\phi_1^\dagger\\
\label{pa10} &&  \frac{\hbar^2}{2m}\nabla^2\phi_1=\big(U_0+ i\overline E \big)\phi_1+U_1\phi_0^\dagger,
\end{eqnarray}
and the complex functions unavoidably equate to
\begin{equation}\label{pa12}
 \phi_0=A_0\exp\Big[\bm{K\cdot x}\Big] \qquad\mbox{and}\qquad \phi_1=A_1\exp\Big[\overline{\bm{K}}\bm{\cdot x}\Big]
\end{equation}
where  $A_0$ and $A_1$ are complex constants, and the complex constant vector $\bm K$ comply with (\ref{pa18}).
Inevitably, (\ref{pa08}-\ref{pa09}) and (\ref{pa12}) implicate the wave function (\ref{pa02}) to be
\begin{equation}\label{pa24}
 \Psi=\mathcal A\exp\left[\bm{K\cdot x}-\frac{E}{\hbar}t\right],
\end{equation}
where the quaternionic amplitude $\mathcal A$ comprises
\begin{equation}\label{pa55}
 \mathcal A=A_0+A_1j,
\end{equation}
and $A_0$ and $A_1$ are of course complex. Taking the conjugate of the second equation in (\ref{pa10}), the spatial functions (\ref{pa12}) implicate the matrix equation
\begin{equation}\label{pa23}
\left[
 \begin{array}{cc}
   U_0+i E & -\, U_1\\
  \overline{ U}_1 & \overline{ U}_0- i\, E
  \end{array}
\right]\left[
\begin{array}{c}
 A_0 \\
 \overline A_1
\end{array}
\right]
=
\frac{\hbar^2}{2m}\bm{K\cdot K}
\left[
\begin{array}{c}
 A_0 \\
 \overline A_1
\end{array}
\right],
\end{equation}
where
\begin{eqnarray}
 \bm{K\cdot K}=\|\bm K_0\|^2-\|\bm K_1\|^2+2i\,\bm{ K_0\cdot K_1}.
\end{eqnarray}

The  characteristic polynomial of the matrix equation (\ref{pa23}) says
\begin{equation}\label{pa11}
\left( U_1+i E-\frac{\hbar^2}{2m}\bm{K\cdot K}\right)\left(\overline U_1- i E-\frac{\hbar^2}{2m}\bm{K\cdot K}\right)+ U_1\,\overline U_1=0.
\end{equation}
Taking the definitions of $U_0$ and $E$ from (\ref{pa15}) and (\ref{pa39}), the real part of (\ref{pa11}) corresponds to
\begin{equation}\label{pa13}
\left[ V_0-\frac{\hbar^2}{2m}\Big(\|\bm K_0\|^2-\|\bm K_1\|^2\Big)\right]^2- E_1^2+\Big( E_0+ V_1\Big)^2-\left(\frac{\hbar^2}{m}\bm K_0\bm{\cdot K}_1\right)^2+ U_1\overline U_1=0,
\end{equation}
and accordingly the imaginary part complies with
\begin{equation}\label{pa14}
E_1\Big( E_0+ V_1\Big)-\left[ V_0-\frac{\hbar^2}{2m}\Big(\|\bm K_0\|^2-\|\bm K_1\|^2\Big)\right]\left(\frac{\hbar^2}{m}\bm K_0\bm{\cdot K}_1\right)=0.
\end{equation}
Engaging (\ref{pa14}) to isolate the real components of $\bm{K\cdot K}$ in (\ref{pa13}), and consequently to eliminate $\,V_0-\frac{\hbar^2}{2m}\Big(\|\bm K_0\|^2-\|\bm K_1\|^2\Big)\,$, one obtains
\begin{equation}
\left[ E_1^2+\left(\frac{\hbar^2}{m}\bm K_0\bm{\cdot K}_1\right)^2\right]\left[\Big( E_0+ V_1\Big)^2 -\left(\frac{\hbar^2}{m}\bm K_0\bm{\cdot K}_1\right)^2\right]+\left(\frac{\hbar^2}{m}\bm K_0\bm{\cdot K}_1\right)^2 U_1\overline U_1=0.
\end{equation}
Moreover, after eliminating $\,2\bm K_0\bm{\cdot K}_1,\,$ (\ref{pa13}) turns into
\begin{multline}
 \left(\left[ V_0-\frac{\hbar^2}{2m}\Big(\|\bm K_0\|^2-\|\bm K_1\|^2\Big)\right]^2-E_1^2\right)\left(\left[ V_0-\frac{\hbar^2}{2m}\Big(\|\bm K_0\|^2-\|\bm K_1\|^2\Big)\right]^2+\Big( E_0+ V_1\Big)^2\right)+\\+\left[ V_0-\frac{\hbar^2}{2m}\Big(\|\bm K_0\|^2-\|\bm K_1\|^2\Big)\right]^2 U_1\overline U_1=0.
\end{multline}
As a result, one determines the real quantities
\begin{equation}\label{uni70}
 \left[ V_0-\frac{\hbar^2}{2m}\Big(\|\bm K_0\|^2-\|\bm K_1\|^2\Big)\right]^2=\frac{\sqrt{\alpha^2+\beta^2}-\alpha}{2}
\end{equation}
and
\begin{equation}
 \left(\frac{\hbar^2}{m}\bm K_0\bm{\cdot K}_1\right)^2=\frac{\sqrt{\alpha^2+\beta^2}+\alpha}{2},
\end{equation}
where one defined
\begin{equation}\label{pa45}
 \alpha=\Big(E_0+ V_1\Big)^2- E_1^2+ U_1\overline{U}_1\qquad \mbox{and}\qquad \beta=2 E_1\big( E_0+ V_1\big)
\end{equation}
in order to finally reach
\begin{equation}\label{pa20}
 \|\bm K_0\|^2=\frac{m}{\hbar^2}\left[V_0\pm\sqrt{\frac{\sqrt{\alpha^2+\beta^2}-\alpha}{2}}+\sqrt{\left(V_0\pm\sqrt{\frac{\sqrt{\alpha^2+\beta^2}-\alpha}{2}}\,\right)^2+\frac{\sqrt{\alpha^2+\beta^2}+\alpha}{2\cos\Omega_0}}\;\right ]
\end{equation}
and 
\begin{equation}\label{pa21}
 \|\bm K_1\|^2=\frac{m}{\hbar^2}\left[-V_0\mp\sqrt{\frac{\sqrt{\alpha^2+\beta^2}-\alpha}{2}}+\sqrt{\left(V_0\pm\sqrt{\frac{\sqrt{\alpha^2+\beta^2}-\alpha}{2}}\,\right)^2+\frac{\sqrt{\alpha^2+\beta^2}+\alpha}{2\cos\Omega_0}}\;\right ].
\end{equation}
Of course, the phase angle defined in (\ref{uni18}) is such that $\cos\Omega_0\neq 0$. Conclusively, from (\ref{pa23}) one obtains
\begin{equation}
A_1=Y_0\overline A_0
\end{equation}
where 
\begin{equation}\label{pa43}
 Y_0=\frac{1}{\overline U_1}\left[-E_1\pm \sqrt{\frac{\sqrt{\alpha^2+\beta^2}-\alpha}{2}}-i\left(E_0+V_1\pm\sqrt{\frac{\sqrt{\alpha^2+\beta^2}+\alpha}{2\cos\Omega_0}}\right)\right],
\end{equation}
and consequently the 
and the solution of the autonomous self-interacting quaternionic particle is thus completed. One only has to notice that the plus signal in (\ref{pa43}) corresponds to $V_0>0$, and changing the signal of this potential accordingly flips the signal.

The physical characterization of the self-interacting autonomous particle permits one to observe the physical expectation values to reproduce the complex particle results (\ref{pa32}), except because of the replacement of the squared amplitude factor, such as
\begin{eqnarray}\label{pa46}
\nonumber&& \left\langle \widehat E\right\rangle=E_1\left(|A_0|^2-|A_1|^2\right)\int\!\!\rho \,dx,\\
\nonumber && \Big\langle \widehat{\bm p}\Big\rangle=\hbar \bm K_1\left(|A_0|^2-|A_1|^2\right) \int\!\!\rho\, dx,\\
&& \left\langle \|\widehat{\bm p}\|^2\right\rangle=\hbar^2\Big(\|\bm K_1\|^2-\|\bm K_0\|^2\Big) \left(|A_0|^2+|A_1|^2\right)\int\!\!\rho\, dx,\\
\nonumber&&\left\langle \widehat V\right\rangle=V_0\left(|A_0|^2+|A_1|^2\right)\int\!\!\rho\, dx.
\end{eqnarray}
with the probability density equal to 
\begin{equation}
\rho=\exp\left[2K_0x-\frac{2E_0}{\hbar}t\right].
\end{equation}
Thus, the conservation of the energy expectation value depends on the wave amplitudes $A_0$ and $A_1$, whose ratio depends on the interaction potential $U_1$ according to (\ref{pa43}). It is indispensable to notice the conformity between the above results to the non-self-interacting case (\ref{pa44}) if $\mathcal E=\overline E$, emphasizing the single difference concerning the ratio between the amplitudes $A_0$ and $A_1$ determined by (\ref{pa42}), what is absent without the self-interaction. The ratio (\ref{pa43}) between the  amplitude factors does not admit a simple and general form, and each particular situation must be considered separately. In the sequel one determines the conditions for stationary states, and entertains the scattering states. A final remark concerning the difference in the amplitude factors of the expectation values of the energy, squared momentum and scalar potential indicates that a difference may appear in the case of normalizable wave functions, and the effect of this difference must be addressed as a future direction of research.

In analogy to complex particles, stationary particles propagate freely in space and time, and require the complex parameters $E$ and $K$ to be pure imaginary. The free parameters are the energy $E$, and the quaternionic scalar potential $U$, and conversely $K$ depends on them. As already discussed, the real component of $E$ must  be zero in to maintain the particle propagation along the time variable. Moreover, the expressions (\ref{pa20}-\ref{pa21}) enable to determine the conditions of the propagation along the space variable, requiring $K_0=0$ and $K_1\neq 0$, and consequently
\begin{equation}
 V_0\pm\sqrt{\frac{\sqrt{\alpha^2+\beta^2}-\alpha}{2}}< 0.
\end{equation}
and 
\begin{equation}\label{pa22}
 \sqrt{\alpha^2+\beta^2}+\alpha=0.
\end{equation}
Undoubtedly, condition (\ref{pa22}) can be rephrased as
\begin{equation}
\alpha\leq 0,\qquad \mbox{and}\qquad \beta=0.
\end{equation}
Remembering (\ref{pa45}), where $\beta=2 E_1\big( E_0+ V_1\big)$, imposing $E_0=0$ and $E_1\neq 0$ for stationary time propagation, and choosing $V_0>0$, one obtains
\begin{equation}
\|\bm K_0\|^2=0,\qquad \mbox{and}\qquad \|\bm K_1\|^2=\frac{2m}{\hbar^2}\left(\sqrt{E_1^2-U_1\overline U_1}-V_0\right),
\end{equation}
what reveals the linear momentum parameter to be decreased when compared to the complex case, and also demonstrates the condition for propagating quaternionic particles to be
\begin{equation}
 E_0=V_1=0,\qquad \mbox{and}\qquad E_1^2>V_0^2+U_1\overline U_1.
\end{equation}
Finally, the complex wave amplitudes $A_0$ and $A_1$ relate as
\begin{equation}
 A_1=\frac{E_1}{\overline U_1}\left(\sqrt{1-\frac{U_1\overline U_1}{E_1^2}}-1\right)\overline A_0.
\end{equation}
Considering that $\sqrt{1-x^2}-1<x$ for $|x|\leq 1$, one concludes that $|A_1|<|A_0|$, and the greater the energy parameter $E_1^2$ in relation to the self-interacting potential $U_1$, the lower the amplitude of the pure quaternionic component of the wave function. Consequently, the self-interaction decreases the contribution of the kinetic energy compared to the participation of the potential energy to the total energy of the particle. 

\paragraph{SCATTERING OF SELF-INTERACTING PARTICLES}  In this one-dimensional situation, the complex potential (\ref{pa40}) is replaced with a quaternionic potential $U$, so that
\begin{equation}
 U=\left\{
 \begin{array}{lll}
 U_I & \mbox{if} & x<0\\
 U_{II} & \mbox{if} & x\geq 0,
 \end{array}
 \right.
\end{equation}
where the quaternionic constants $U_I$ and $U_{II}$ conform (\ref{pa25}-\ref{pa15}). The wave function that describes the scattering phenomenon between the regions governed respectively by  $U_I$, and $U_{II}$, following (\ref{pa24}) accordingly comprises
\begin{equation}
 \Psi=\left\{ 
 \begin{array}{lll}
  \Psi_I=\Big(1+H_0j\Big)\Big(\exp\big[K_Ix\big]+ R\exp\big[- K_Ix\big]\Big)\exp\left[-\frac{E}{\hbar}t\right] & \mbox{para} & x<0\\
  \Psi_{II}=\Big(1+I_0j\Big) T\exp\left[K_{II}x-\frac{E}{\hbar}t\right] & \mbox{para} & x\geq 0,
 \end{array}
 \right.
\end{equation}
where $R$ and $T$ are complex constants, and $H_0$ and $I_0$ are complex components of the quaternionic amplitude that follow (\ref{pa43}). The solution is analgous to the complex case, but with the additional constraint
\begin{equation}
H_0=I_0.
\end{equation}
Moreover, one cannot forget the $u$ parameter on (\ref{pa38}), that also depends on the components of $K_I$ and $K_{II}$, and therefore the transmission rates differ from the complex case. Consequently, the self-interaction solution is more constrained than the previous complex solution, although it is qualitatively similar, and thus complying with an expectation, because one does not expect a quaternionic particle to be a completely different physical object compared to a complex particle, but solely something where additional possibilities can be found. On the other hand, the precise effects of each parameter on the solution, and the possible physical interpretation of these fields are interesting directions for future research.

\section{QUATERNIONIC PARTICLES II}
In this section, one considers the right quaternionic wave equation, that is the remaining alternative to (\ref{pa04}), and  of course reads
\begin{equation}\label{pa05}
 \hbar\frac{\partial\Psi}{\partial t}i=\widehat{\mathcal H}\Psi.
\end{equation}
and consequently the wave function (\ref{pa02}) produces the complex system of equations
\begin{eqnarray}
&&\;\;\; i\hbar\frac{\partial\psi_0}{\partial t}=-\frac{\hbar^2}{2m}\nabla^2\psi_0+U_0\psi_0-U_1\psi_1^\dagger\\
&& - i\hbar\frac{\partial\psi_1}{\partial t}=-\frac{\hbar^2}{2m}\nabla^2\psi_1+U_0\psi_1+U_1\psi_0^\dagger.
\end{eqnarray}
Constant scalar potentials $U_0$ and $U_1$ leads to
\begin{equation}
\left[
 \begin{array}{cc}
   U_0+i E & -\, U_1\\
  \overline U_1 & \overline U_0+i\,E
  \end{array}
\right]\left[
\begin{array}{c}
 A_0 \\
 \overline A_1
\end{array}
\right]
=
\frac{\hbar^2}{2m}\bm{K\cdot K}
\left[
\begin{array}{c}
 A_0 \\
 \overline A_1
\end{array}
\right],
\end{equation}
Accordingly, the real part of the characteristic polynomial comprises,
\begin{equation}\label{pa50}
\left[V_0- E_1-\frac{\hbar^2}{2m}\Big(\|\bm K_0\|^2-\|\bm K_1\|^2\Big)\right]^2+ V_1^2-\left( E_0-\frac{\hbar^2}{m}\bm K_0\bm{\cdot K}_1\right)^2+ U_1\overline U_1=0,
\end{equation}
and the imaginary part inevitably reads
\begin{equation}\label{uni91}
\left[V_0- E_1-\frac{\hbar^2}{2m}\Big(\|\bm K_0\|^2-\|\bm K_1\|^2\Big)\right]\left( E_0-\frac{\hbar^2}{m}\bm K_0\bm{\cdot K}_1\right)=0.
\end{equation}
Non-trivial solutions to the above system require that 
\begin{equation}
  E_0-\frac{\hbar^2}{m}\bm K_0\bm{\cdot K}_1\neq 0
\end{equation}
as otherwise only non-self-interaction solutions hold. Therefore, 
\begin{equation}
 V_0- E_1-\frac{\hbar^2}{2m}\Big(\|\bm K_0\|^2-\|\bm K_1\|^2\Big)=0,
\end{equation}
a relation that must be valid in the self-interacting case as well as in the non-self-interacting case. Therefore, one obtains
\begin{eqnarray}
 && \|\bm K_0\|^2=\frac{m}{\hbar^2}\left[V_0-E_1+\sqrt{\big(V_0-E_1\big)^2+\left(\frac{E_0\pm\sqrt{V_1^2+U_1\overline U_1}}{\cos\Omega_0}\right)^2}\,\right]\\
 && \|\bm K_1\|^2=\frac{m}{\hbar^2}\left[E_1-V_0+\sqrt{\big(V_0-E_1\big)^2+\left(\frac{E_0\pm\sqrt{V_1^2+U_1\overline U_1}}{\cos\Omega_0}\right)^2}\,\right]
\end{eqnarray}
The above equations indicate that pure stationary states are not viable solutions of (\ref{pa05}) within the coupled self-interacting regime.  The analysis of the scattering case follows the previous quaternionic case, and seems not deserving of any further analysis.

\section{CONCLUSION\label{F}}%

This article describes relevant features of the real Hilbert space formalism of $\mathbbm H$QM. First of all, it permits the analysis of the energy conservation of non-stationary processes, something that $\mathbbm C$QM is unable to obtain. The results also permit a precise physical interpretation of each component of the quaternionic scalar potential, something that was never reached in the anti-hermitean formalism of $\mathbbm H$QM. Besides, it determines the quaternionic components of the scalar potential to support the self-interaction between the complex components of the quaternionic quantum particles. 

In summary, the novel results contained in this article for the autonomous particle can be applied to several more sophisticated physical models, where the self-interaction has never been considered. The directions of future research are consequently various, what ascribes potential importance to the results presented in this article.

\paragraph{DATA AVAILABILITY STATEMENT} The author declares that data sharing is not applicable to this article as no data sets were generated or analysed during the current study.

\paragraph{DECLARATION OF INTEREST STATEMENT} The author declares that he has no known competing financial interests or personal relationships that could have appeared to influence the work reported in this paper.

\paragraph{FUNDING} This work is supported by the Funda\c c\~ao de Amparo \`a Pesquisa do Rio Grande do Sul, FAPERGS, grant 23/2551-0000935-8 within edital 14/2022.

%
%
%
%

\begin{footnotesize}
\bibliographystyle{unsrt} 
\bibliography{bib_qlivre}
\end{footnotesize}
\end{document}